\documentclass{iaus}
\usepackage{graphicx}

\title[JD 11.~~The Light-Time Effect in the eclipsing binaries GK Cep and VY Cet] 
{The Light-Time Effect in the eclipsing \\ binaries GK Cep and VY Cet}

\author[A. Liakos \& P. Niarchos]   
{Alexios Liakos
 \and Panagiotis Niarchos}

\affiliation{Department of Astrophysics,  Astronomy and Mechanics,  Faculty of Physics,
\\National and Kapodistrian University of Athens,  Athens, Greece
 \\ email: {\tt alliakos@phys.uoa.gr, pniarcho@phys.uoa.gr } \\[\affilskip]}

\pubyear{2008}
\volume{xxx}  
\pagerange{119--126}
\jname{Title of your IAU Symposium}
\editors{A.C. Editor, B.D. Editor \& C.E. Editor, eds.}
\begin{document}

\maketitle

\begin{abstract}
New times of minima of the eclipsing binaries GK Cep and VY Cet, obtained at the  Observatory of the University of Athens, have been used together with all reliable timings found in the literature in order to study the period variation and search for the presence of third body in the systems.
\keywords{GK Cep, VY Cet, Light Time Effect, eclipsing binaries, period variation, O-C study.}
\end{abstract}

\firstsection 

\section{Observations of the systems and analysis of the O-C diagrams}

Both systems were observed with the 40-cm Cassegrain telescope of the Observatory of the University of Athens, equipped with the ST-8XMEI CCD camera and Bessell VRI filters. For GK Cep the observations were carried out during one night in January 2007 and two nights in June 2007, and for VY Cet during one night in December 2006 and one night in January 2007. For GK Cep the stars  SAO 10066 and  GSC 4465:0372, and for VY Cet the stars GSC 5857:1891 and GSC 5857:1297 were used as  comparison and check stars, respectively. Two primary and one secondary minima for GK Cep, and one primary and one secondary minima for VY Cet were obtained and they are given in Table  \ref{tab1}.

\begin{figure}
\begin{center}
 \includegraphics[width=5.0in]{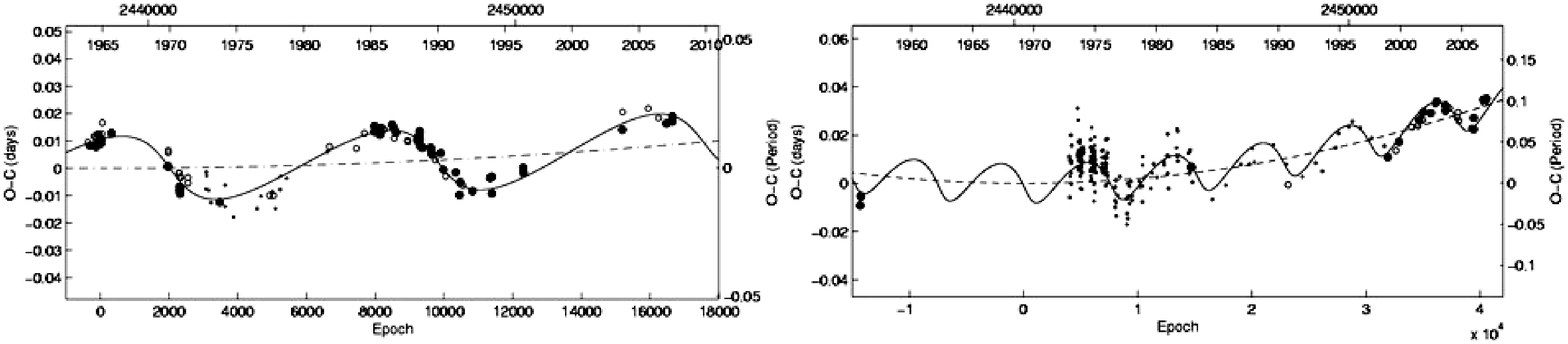}
 \caption{The O-C diagrams of GK Cep (left) and VY Cet (right) with the theoretical fittings.}
   \label{fig1}
\end{center}
\end{figure}

We used only reliable times of minima, 120 for GK Cep and 231 for VY Cet, taken from the literature, and the new minimum times from our observations. The O-C diagrams have been analyzed with the least-squares method and the results are presented in the Figure \ref{fig1}. Figure \ref{fig1} shows the O-C diagrams fitted by a parabolic and sinusoidal curve, thus suggesting a mass transfer between the components and the presence of a third body in the systems. The new elements for the binaries, the orbital parameters of possible third bodies and their physical parameters, using the assumption that they are MS stars ($L_{3}=M_{3}^{3.5}$),  are given in Table  \ref{tab2}.

\begin{table}
  \begin{center}
  \caption{The obtained times  of minima }
  \label{tab1}
\begin{tabular}{|c|c|c|c|}\hline
   System & HJD (2450000.0+) & Error & Type \\\hline
   GK Cep & 54114.27388 & 0.00002 & I \\
   GK Cep & 54273.42250 & 0.00010 & I \\
   GK Cep & 54279.50782 & 0.00013 & II \\
   VY Cet & 54090.23533 & 0.00019 & II \\
   VY Cet & 54102.33443 & 0.00008 & I \\\hline

 \end{tabular}
 \end{center}

 \end{table}

\begin{table}
  \begin{center}
  \caption{The new parameters of GK Cep and VY Cet and the orbital and physical parameters of the third body in each system.}
  \label{tab2}
 {\scriptsize

 \begin{tabular}{|c|c|c|}\hline
   parameter & GK Cep & VY Cet \\ \hline
   \textbf{The orbital and physical parameters of the EB}  \\ \hline
   $M_{1}+M_{2} (M_{\bigodot})$ & 2.7 + 2.5 & 1.02 + 0.68 \\
   Sp. Type & A2V + A2V & G5 \\
   q (days/cycle) ($\ast10^{-10})$ & 0.30960 (4) & 0.19720 (1) \\
   $\dot{P}$ (days/yr) ($\ast10^{-8})$ & 2.4158 (3) & 4.2267 (2) \\
   $\dot{M}$ ($M_{\bigodot}$/yr) ($\ast10^{-7})$  & 2.9031 (3) & 8.4333 (4) \\
   A (semi-amplitude of O-C) (days) & 0.0117 (4) & 0.0082 (5) \\
   $x^{2}$ & 0.0063854895 & 0.0103336605 \\ \hline
   \textbf{The orbital parameters of the 3rd body}\\ \hline
   $e_{3}$/$\omega_{3}$& 0.45 (6)/181.4 (6.5) & 0.39 (13)/219.0 (18.5)  \\
   T (periastron passage) (HJD) & 2455140.816 (155.377) & 2430113.271 (186.074) \\
   f($m_{3}$)$(M_{\bigodot})$ & 0.02994 (1) & 0.0649 (1) \\
   $P_{3}$ (yrs) & 19.91 (0.19) & 7.07 (0.06) \\ \hline
   \textbf{The physical parameters of the 3rd body}\\ \hline
      $m_{3,min}(M_{\bigodot}) (i=90^{0})$& 1.0641 (3) & 0.7250 (9) \\
   Sp. Type $^{*}$/ $T_{3} (K)^{*}$& G0V/ 5940 & K3V/ 5940\\
   $L_{3}(L_{\bigodot})^{*}$/$R_{3}(R_{\bigodot})^{*}$/$M_{3}(mag)^{*}$ & 1.35/1.1/4.4 & 0.26/0.77/6.7 \\\hline

 \end{tabular}

  }
 \end{center}
\vspace{0.5mm}
 \scriptsize{
 {\it$^{\ast}$assumed}\\
  }
\end{table}

\section{Discussion and conclusions}

New orbital and physical parameters for the eclipsing binaries GK Cep and VY Cet have been derived by means of an analysis of their O-C diagrams. A third body in an eccentric orbit
was found for both systems. The period and minimum mass of these bodies are given in Table 2.
The light contribution of the third body to the total light was found to be 2.1\% for GK Cep, too small to be detected by any method, while that for the VY Cet was
found to be about 19.6\%, large enough and detectable by observational methods.
Quadratic ephemerides T = 2438694.69719(159) + $0.9361642(2)\times E + 0.30960(4) \times 10^{-10} \times E^{2}$ for GK Cep; and T = 2440282.14835(132) + 0.3408091(1) $\times E + 0.19720(1) \times 10^{-10} \times E^{2} $ for VY Cet, were also found. For both cases the parabolic term indicates a mass transfer from the less massive to the more massive component of the system.

\section{Acknowledgements}

This work has been financially supported by the Special Account for Research Grant No 70/3/8680 of the National and Kapodistrian University of Athens, Greece.

\end{document}